\documentclass{ws-procs975x65}            
\begin{document}
\title{Irreversible matter creation processes through a nonminimal
curvature-matter coupling}

\author{Francisco S. N. Lobo$^1$, Tiberiu Harko$^2$, Jos\'e P. Mimoso$^1$, and Diego Pav\'{o}n$^3$}

\address{$^1$Instituto de Astrof\'{\i}sica e Ci\^{e}ncias do Espa\c{c}o \& Departamento de F\'{\i}sica, Faculdade de
Ci\^encias da Universidade de Lisboa, Edif\'{\i}cio C8, Campo Grande,
P-1749-016 Lisbon, Portugal\\
E-mail: fslobo@fc.ul.pt; jpmimoso@fc.ul.pt}

\address{$^2$Department of Mathematics, University College London, Gower Street, London
WC1E 6BT, United Kingdom. E-mail: t.harko@ucl.ac.uk}

\address{$^3$Departamento de F\'{\i}sica, Universidad Aut\'{o}noma de Barcelona, 08193
Bellaterra (Barcelona), Spain.
E-mail: diego.pavon@uab.es}

\begin{abstract}
An interesting cosmological history was proposed by
Prigogine {\it et al.} who considered the Universe as a
thermodynamically open system. This scenario is characterized by
a process of matter creation, which corresponds to an irreversible
energy flow from the gravitational field to the pressureless
matter fluid. Here, we show that the gravitationally induced
particle production may arise from a nonminimal curvature-matter coupling. 
By considering the equivalent scalar-tensor theory, the
cosmological implications of the model are discussed. 
As all known natural systems tend to a state of thermodynamic equilibrium, and
assuming the universe is not different in this respect, we also
discuss the conditions to attain the equilibrium state.
\end{abstract}


\bodymatter

\section{Introduction}

An interesting type of cosmological history, including
large-scale entropy production, was proposed by Prigogine {\it et
al.} \cite{Pri0,Pri}, who considered the cosmological thermodynamics of
open systems. The role of irreversible processes corresponding to
matter creation in General Relativity was investigated, and the
usual adiabatic energy conservation laws were modified. The corresponding
cosmologies were based on a reinterpretation of the matter
stress-energy tensor in Einstein's equations, with the matter
creation corresponding to an irreversible energy flow from the
gravitational field to the created matter fluid \cite{Cal}.
It was shown that the 2nd law of thermodynamics allows
spacetime to transform into matter, forbidding the inverse
process. It appears that the conventional initial
big bang singularity is structurally unstable with respect to
irreversible matter creation. Thus, the cosmological
history in the framework of the thermodynamics of open systems
starts from an instability of the quantum vacuum rather than from
a singularity. A remarkable fact is that the de Sitter stage
appears to be an attractor independently of the initial
fluctuation \cite{Pri0,Pri}.

The final de Sitter phase is particularly relevant due
to the recent discovery of the late-time cosmic acceleration
\cite{1,2}, where the simplest explanation is to invoke a
cosmological constant. The latter is usually
associated to the quantum vacuum energy, which may not be
constant but decays into radiation and matter particles.
In this work, we consider an alternative mechanism for the
gravitational particle production, namely, in modified gravity,
through a nonminimal curvature-matter coupling
\cite{Bertolami:2007gv}. A general property of these theories is
the non-conservation of the energy-momentum tensor
\cite{Bertolami:2007gv,Harko:2008qz,Harko:2010hw,Harko:2011kv}.
Thus, the coupling between the matter and the higher derivative
curvature terms may be interpreted as an exchange of energy and
momentum between both, and we argue that  this mechanism naturally
induces a gravitational particle production \cite{Harko:2015pma}.

\section{Thermodynamic open systems and curvature-matter couplings}

We consider the spatially-flat
Friedmann-Robertson-Walker (FRW) metric, 
$ds^2=dt^2-a^2(t)\left(dx^2+dy^2+dz^2\right)$, where $a(t)$
denotes the scale factor. Using the thermodynamics of open systems, the number of particles is not conserved, i.e., the usual
expression for the evolution of the number density, $\dot{n} \, +
\, 3 H n = 0$, where $H = \dot{a}/a$, generalizes to $\dot{n}+3nH=\Gamma n$.
Here, the quantity $\Gamma$ stands for the particle creation rate,
which  as imposed by the second law of thermodynamics obeys
$\Gamma \geq 0$. Therefore, the energy conservation equation
describing an irreversible particle creation can be rewritten as an effective energy
conservation equation,
\begin{equation}
\dot{\rho}+3\left( \rho +p+p_{c}\right) H=0, \qquad {\rm with} \qquad p_{c} = -\frac{\rho +p}{3}\frac{\Gamma }{H}. 
  \label{comp}
\end{equation}%
where $p_{c}$ is a new thermodynamic quantity, $p_{c}$, denoted the
creation pressure \cite{Pri}.

In the following, we argue that an explicit coupling between curvature
and matter generates an irreversible energy flow from the gravitational
field to newly created matter constituents, with the second law of
thermodynamics requiring that space-time transforms into matter.
The gravitational theory, with a nonminimal coupling between matter and curvature\cite{Bertolami:2007gv}, is given by the Lagrangian $L=\frac{1}{2}f_1(R)+\left[1+\lambda f_2(R)\right]{\ L}_{m}$,
where the factors $f_i(R)$ (with $i=1,2$) are arbitrary functions of the
Ricci scalar $R$. The coupling constant $\lambda$ determines the strength of
the interaction between $f_2(R)$ and the matter Lagrangian density, ${L}_{m}$.

The curvature-matter coupling is equivalent to a
two-potential scalar-tensor Brans-Dicke type theory
\cite{Faraoni:2007sn}, with the following action
\begin{equation}  \label{300}
S=\int  \left[ \frac{\psi R }{2} -V(\psi)\, +U(\psi) L_m \right]\,
\sqrt{-g} \, d^4x  \, ,
\end{equation}
where the two potentials $V(\psi)$ and $U(\psi)$  are defined as
\begin{equation}  \label{400}
V(\psi) = \frac{\phi(\psi) f_1^{\prime }\left[ \phi (\psi ) \right] -f_1%
\left[ \phi( \psi ) \right] }{2}, \qquad U( \psi) = 1+\lambda
f_2\left[ \phi( \psi ) \right]\,.
\end{equation}

By varying the action (\ref{300}) with respect to
$g_{\mu \nu}$ and to the field $\psi$ provides the gravitational
field equations
\begin{equation}  \label{14}
\hspace{-0.1cm}\psi \left(R_{\mu \nu}-\frac{1}{2}g_{\mu \nu}R\right)+
\left(g_{\mu \nu }\nabla _{\alpha}\nabla^{\alpha} -\nabla _{\mu }\nabla _{\nu }\right)\psi =
U(\psi)T_{\mu \nu}+V(\psi)g_{\mu \nu},
\end{equation}
\begin{equation}  \label{15}
\frac{R}{2}-V^{\prime }(\psi)+U^{\prime }(\psi)L_m=0.
\end{equation}

The divergence of the energy-momentum tensor provides the following energy balance equation
\begin{equation}  \label{eneq}
\dot{\rho}+3H(\rho +p)+\rho \frac{d}{ds}\ln U(\psi )+\frac{2V^{\prime
}\left( \psi \right) -U^{\prime }(\psi )L_{m}}{U(\psi )}\dot{\psi}=0\,.
\end{equation}

Therefore, in modified gravity with a linear
curvature-matter coupling, and confronting with Eq. (\ref{comp}),
the creation pressure is given by
\begin{equation}  \label{pc}
p_{c}=-\frac{1}{3H}\left\{\rho \frac{d}{dt}\ln U(\psi )+\frac{2V^{\prime
}\left( \psi \right) -U^{\prime }(\psi )L_{m}}{U(\psi )}\dot{\psi}\right\},
\end{equation}
where the particle creation rate $\Gamma $ is  a non-negative quantity defined as
\begin{equation}  \label{33}
\Gamma =-\frac{1}{\rho +p}\Bigg\{\rho \frac{d}{dt}\ln U(\psi )+\frac{%
2V^{\prime }\left( \psi \right) -U^{\prime }(\psi )L_{m}}{U(\psi )}\dot{\psi}%
\Bigg\}.
\end{equation}
Note that from Eq. (\ref{eneq}), the change in the particle number is due to the
transfer of energy-momentum from gravity to matter.

In the present work, we define the entropy through the
particle production rate, as depending on the positive particle
creation rate $\Gamma $ via the relation $\dot{S}/S=\Gamma \geq
0$. Therefore, in an ever expanding Universe with particle
creation, the matter entropy will increase indefinitely. On the
other hand, natural systems, left to themselves, tend to attain a
state of thermodynamic equilibrium. This implies that the entropy
of isolated systems never decreases, $\dot{S}\geq 0$, and that it
is concave, at least in the last stage of  approaching
thermodynamic equilibrium.

The validity of the second law of thermodynamics in
cosmology was investigated in detail in \cite{Mim}. There
it was shown that if one defines the entropy, $S$, measured by a
comoving observer as the entropy of the apparent horizon plus that
of matter and radiation inside it, then the Universe approaches
thermodynamic equilibrium as it nears the final de Sitter phase.
Then it follows that $S$ increases, and that it is concave, thus
leading to the result that the second law of thermodynamics is
valid, as one should expect given the strong connection between
gravity and thermodynamics, for the case of the expanding
Universe.

In the case of a spatially-flat FRW Universe filled with dust, one
has
\begin{equation}
S = S_{ah} + S_m = \frac{\pi}{H^2} + \frac{4\pi}{3H^3}\,n(t)\;,
\label{MP}
\end{equation}
where we have used the fact that the radius of the apparent horizon is $%
r_{ah}= H^{-1}$ \cite{Bak}.
Therefore, the thermodynamic requirements
$S^{\prime}\geq 0 $ and $S^{\prime \prime }\leq 0$ impose specific
constraints on the particle creation rate $\Gamma $, and its
derivative with respect to the scale factor. A particularly
important case is that of the de Sitter evolution of the Universe,
with $H=H_{\star}=\mathrm{constant}$. In this case, we have
\begin{equation}
S^{\prime }= \frac{4\pi }{ 3H_{\star}^{4}}\frac{\left[ \Gamma (a)
-3H_{\star} \right]}{a}n(a) \geq 0  \,,
\end{equation}
\begin{eqnarray}
S^{\prime \prime}= \frac{4 \pi n(a)}{3 a^2 H_{\star}^5}
\Big\{\Gamma ^2(a)+H_{\star} \left[a \Gamma ^{\prime }(a)+12
H_{\star}\right] -7 H_{\star} \Gamma (a)\Big\} \leq 0 \,.
\end{eqnarray}
Accordingly, the constraints $\Gamma \geq 3H_{\star}$ and $\Gamma ^{\prime
}(a)\leq \left[7\Gamma (a)-\Gamma
^2(a)/H_{\star}-12H_{\star}\right]/a$ are imposed on the particle creation
rate $\Gamma$. For $\Gamma =3H_{\star}$, we obtain $S =
\mathrm{constant}$, showing that in this case the cosmological
evolution is isentropic. Here, $H_{\star}$ denotes the expansion
rate of the final de Sitter phase. Realistically, it must be lower
than the Hubble constant value, $H_{0} \simeq 73.8$km/s/Mpc.

\section{Conclusion}

In conclusion, the gravitational theory investigated in
the present essay suggests that the curvature-matter coupling may well
be a leading feature of our Universe.  The current accelerated
cosmic expansion  may be seen as a hint for matter creation which
offers a welcome alternative to the elusive dark energy. The
existence of some forms of the curvature-matter coupling leading
to matter creation processes is not ruled out by the available
data. Presumably, the functional forms of the potentials $V(\psi)$
and $U(\psi)$ that fully characterize the nonminimal
curvature-matter coupling gravitational theory may be provided by
fundamental quantum field theoretical models of the gravitational
interaction, thus opening the possibility to an in-depth comparison
of the predictions of the gravitational theory with cosmological
and astrophysical data.

\section*{Acknowledgments}
FSNL was supported by a FCT Research contract, with
reference IF/00859/2012, funded by FCT/MCTES (Portugal). FSNL and
JPM acknowledge financial support of the Funda\c{c}\~{a}o para a
Ci\^{e}ncia e Tecnologia through the grant EXPL/FIS-AST/1608/2013
and UID/FIS/04434/2013. DP was partially supported by the
``Ministerio de Econom\'{\i}a y Competitividad, Direcci\'{o}n
General de Investigaci\'{o}n  Cient\'{\i}fica y T\'{e}cnica",
Grant N$_{o}$. FIS2012-32099.

\bibliographystyle{ws-procs975x65}
\bibliography{ws-pro-sample}



\end{document}